\documentclass[sigplan,10pt,review,anonymous]{acmart}
\acmConference[PPoPP'24]{29th ACM SIGPLAN Annual Symposium on Principles and Practice of Parallel Programming}{02 - 06 March, 2024}{Edinburgh}
\acmYear{2024}
\acmISBN{} 
\acmDOI{} 
\startPage{1}

\setcopyright{none}

\usepackage{booktabs}   
\usepackage{subcaption} 

\definecolor{darkGreen}{HTML}{38761d}
\definecolor{darkRed}{HTML}{8b0000}
\usepackage{diagbox}

\usepackage[colorinlistoftodos,prependcaption,
textsize=tiny,
color=limeyellow,linecolor=magenta]{todonotes}

\usepackage[normalem]{ulem}
\usepackage[font=small,labelfont=bf, textfont=bf]{caption}
\usepackage{tabularx}
\usepackage{amsmath}

\usepackage{amsmath}
\usepackage{algorithm}
\usepackage{algpseudocode}
\usepackage{soul}
\usepackage{slashbox}
\usepackage[compact]{titlesec}
\usepackage{float}
\usepackage{wrapfig}
\usepackage{multicol}
\usepackage{tabularray}
\usepackage{multirow}
\usepackage{pifont}
\newcommand{\cmark}{\ding{51}}%
\newcommand{\xmark}{\ding{55}}%

\usepackage[most]{tcolorbox} 

\usepackage{enumitem}
\setlist[itemize]{noitemsep, topsep=-\parskip, leftmargin=*}
\setlist[enumerate]{noitemsep, topsep=-\parskip, leftmargin=*}

\newcommand{\ie}{\textit{i}.\textit{e}.,}
\newcommand{\eg}{\textit{e}.\textit{g}.}

\newcommand{\sysname}[1]{\textit{HaX-CoNN}} 

\begin{document}

\title{Shared Memory-contention-aware Concurrent DNN Execution for Diversely Heterogeneous System-on-Chips}


\author{First1 Last1}
\authornote{with author1 note}          
\orcid{nnnn-nnnn-nnnn-nnnn}             
\affiliation{
  \position{Position1}
  \department{Department1}              
  \institution{Institution1}            
  \streetaddress{Street1 Address1}
  \city{City1}
  \state{State1}
  \postcode{Post-Code1}
  \country{Country1}                    
}
\email{first1.last1@inst1.edu}          

\author{First2 Last2}
\authornote{with author2 note}          
\orcid{nnnn-nnnn-nnnn-nnnn}             
\affiliation{
  \position{Position2a}
  \department{Department2a}             
  \institution{Institution2a}           
  \streetaddress{Street2a Address2a}
  \city{City2a}
  \state{State2a}
  \postcode{Post-Code2a}
  \country{Country2a}                   
}
\email{first2.last2@inst2a.com}         
\affiliation{
  \position{Position2b}
  \department{Department2b}             
  \institution{Institution2b}           
  \streetaddress{Street3b Address2b}
  \city{City2b}
  \state{State2b}
  \postcode{Post-Code2b}
  \country{Country2b}                   
}
\email{first2.last2@inst2b.org}         


\begin{abstract}
Two distinguishing features of state-of-the-art mobile and autonomous systems are 1) there are often multiple workloads, mainly deep neural network (DNN) inference, running \textit{concurrently} and \textit{continuously}; and 2) they operate on shared memory system-on-chips (SoC) that embed heterogeneous accelerators tailored for specific operations. 
State-of-the-art lacks efficient performance and resource management techniques necessary to either maximize total system throughput or minimize end-to-end workload latency.
In this work, we propose \sysname{}, a novel scheme that characterizes and maps layers in concurrently executing DNN inference workloads to a diverse set of accelerators within a SoC. 
Our scheme uniquely takes per-layer execution characteristics, shared memory (SM) contention and inter-accelerator transitions into account to find \textit{optimal} schedules. 
We evaluate \sysname{} on NVIDIA Orin, NVIDIA Xavier, and Qualcomm Snapdragon 865 SoCs. Our experimental results indicate that \sysname{} minimize the memory contention by up to 45\% and can improve latency and total throughput by up to 32\% and 29\%, respectively, in comparison to the state-of-the-art approaches.
\end{abstract}



\maketitle

\section{Introduction}
\vspace{-0.4em}





Modern mobile and autonomous systems ---such as cars, drones, and robots--- hinge on edge intelligence, which involves running computationally demanding workloads~\cite{krishnan2022roofline,krishnan2022automatic,wan2022analyzing}. 
Notably, a diverse range of applications embed multiple DNNs as subtasks, such as object detection and semantic segmentation for autonomous systems~\cite{grigorescu2020survey,autopilot_tesla} or pose estimation and eye-tracking for VR applications~\cite{wang2020vr}. Workloads running \textit{concurrently} and \textit{continuously} in such systems necessitates powerful SoCs that can meet high computational demand and ensure safety~\cite{guiochet2017safety} and QoS~\cite{hudson2021qos} requirements. Thus, such SoCs are often equipped with a CPU, a GPU, and one or more domain-specific accelerators (DSAs), optimized to perform specific type of operations (OPs). 
For example, NVIDIA's Xavier AGX and Orin architecture comprises deep learning accelerators (DLA), and programmable vision accelerators (PVA) in addition to the CPU and the GPU. 
Utilizing different type of DSAs for concurrently running workloads 
enables the flexibility to explore execution strategies~\cite{monil2020mephesto}. \textit{Leveraging this opportunity, in this work, we focus on mapping layers of parallel DNNs to different types of DSAs so that we can improve computational latency and system throughput.}




A commodity feature such heterogeneous SoCs have in common is the \textit{shared physical main memory} where the data and accessed by all processing units (PUs), \ie{}, CPUs, GPUs, and DSAs.
Even though this cost-driven design decision curbs costly data movement, memory subsystems in such architectures are often designed to accommodate the memory demands of a single PU in isolation. 
Consequently, \textit{shared memory contention emerges as one of the primary performance bottlenecks in mobile and autonomous SoCs when parallel OPs are mapped to different accelerators}~\cite{luan2022many,Pccs_micro,hill2019gables}.

\begin{figure}[t]
    \centering

    \includegraphics[width=0.85\linewidth]{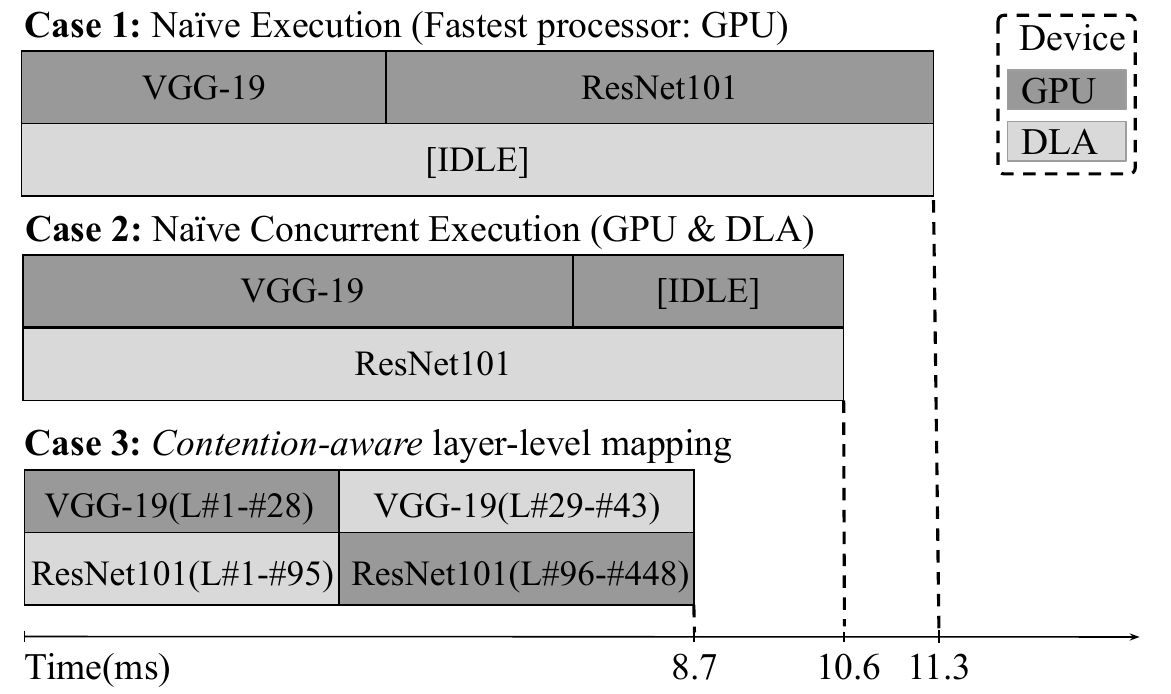}
    \vspace{-1em}
    \caption{Different ways of executing VGG-19 and ResNet-101 DNNs in parallel on Xavier AGX. }
    \label{fig:proposed_methodology}
    \vspace{-1.5em}
\end{figure}

We investigate concurrent execution on shared memory SoCs through a case study. In a typical loop running on autonomous systems, 
VGG-19~\cite{simonyan2015vgg} and ResNet101~\cite{He_2016_CVPR} can be used in tandem for vision, \ie{} perception, tasks. Since the remaining tasks in the autonomous loop depend on the completion of these two perception OPs, utilizing all computational resources in the SoC for these OPs is expected to reduce the total latency of the system. 
Fig. \ref{fig:proposed_methodology} illustrates three different ways of executing these two DNNs on NVIDIA Xavier AGX SoC. 
In \textbf{Case 1}, DNNs are \textit{serially} executed on the fastest DSA (\ie{} GPU), resulting in 11.3ms of cumulative latency. However, this naïve method leaves DLA idle, thereby underutilizing the system resources.
The first approach can be improved by a \textit{naïve concurrent} execution, as shown in \textbf{Case 2}. In this scheme, VGG-19 is run on GPU, and ResNet101 is mapped to DLA, resulting in a slight improvement in cumulative latency: 10.6ms. 
However, the speed-up obtained remains limited due to two reasons: (1) DLA takes longer to execute, leaving GPU idle towards the end, and (2) when GPU and DLA operate together, they contend for shared memory and slow down. \textit{For more efficient execution of concurrent DNNs, we need a finer-granular mapping of the workload, \ie{} layer-level, to DSAs. }

\textbf{Case 3} depicts an ideal case where layers in each DNN are divided into two groups, \eg{} VGG-19 and ResNet101 are divided after layers \#28 and \#95, respectively. For each DNN, the execution is switched between the two DSAs at the boundary of corresponding layer groups (\ie{} transition point). While seemingly non-intuitive, this approach considerably improves the cumulative latency and increases the overall system utilization. This is due to a careful partitioning and mapping of layers to GPU and DLA in a way that: (1) shared memory contention across concurrently running layers is minimized, (2) neither of the DSAs is left idle, and (3) the overhead of switching between accelerators between two layers is minimized. However, finding such partitioning is not trivial and the state-of-the-art (which is detailed in Section~\ref{sec:related_work}) fails to provide a holistic approach to perform this partitioning optimally.


In this study, we propose \sysname{}\footnote{Upon publication, the full implementation of \sysname{} will be provided as an open source framework.}, a \textit{multi-accelerator and contention-aware execution scheme}, for collaboratively and concurrently running DNNs on shared memory SoCs.
\sysname{} is centered around characterizing common layers in DNNs according to their DSA-specific performance and identifying how they are affected by the shared memory contention. Leveraging \textit{decoupled} performance and contention characterization \textit{at a layer-level}, \sysname{} exploits distinct capabilities of each DSA in the system by deciding whether the execution of next layer in the DNN should \textit{transition} to another DSA or not. \sysname{} uniquely finds an \textit{optimal} mapping between the layers and DSAs in the system by formulating the problem as a set of constraint-based linear equations and utilizing SAT solvers to find a solution. 
%

Our work makes the following contributions:
\begin{itemize}
    \item We present \sysname{}, a contention-aware, multi accelerator execution scheme that \textit{maximizes compute utilization }and \textit{minimizes the overall latency} of concurrently running DNNs on shared memory SoCs.  
    \item We propose a generalized and formal layer-to-accelerator mapping approach for concurrently running DNNs. Our work demonstrates that SAT solvers can be utilized to produce \textit{optimal schedules} for \textit{multi-accelerator} execution. 
    \item We build a new contention modeling approach which significantly reduces profiling search space by decoupling performance measurement and the slowdown due shared memory usage.
    \item We present, D-\sysname{}, a dynamic runtime adaptation of SAT solver-based optimal schedule generation for dynamically changing workloads.
    \item We evaluate \sysname{} and D-\sysname{} on NVIDIA AGX Orin, Xavier AGX, and Qualcomm Snapdragon 865 SoCs. Our results show that \sysname{} can provide latency throughput improvements up to 32\% and 29\% respectively, over greedy-scheduling based approaches.
\end{itemize}

\section{Related Work} \label{sec:related_work}

\begin{table}[b]
\vspace{-1.5em}
\scriptsize
\centering
\caption{Comparison of the features offered by the most related work.}
\label{table:related_work}
\vspace{-1em}
\begin{tabular}{ | m{2.3cm}|  m{0.3cm} | m{0.3cm}| m{0.3cm}| m{0.3cm}| m{0.3cm} | m{0.3cm} | m{0.3cm} |  m{0.3cm} |  }
  \hline
  Related Work 
  & \rotatebox{90}{Mensa~\cite{boroumand2021google}}
  & \rotatebox{90}{AxoNN~\cite{dagli22AxoNN}}
  & \rotatebox{90}{Pipeline~\cite{jeong2021deep}}
  & \rotatebox{90}{OmniBoost~\cite{karatzas2023omniboost}}
  & \rotatebox{90}{MoCA~\cite{kim2023moca}}
  & \rotatebox{90}{Herald~\cite{kwon2021hpca}}
  & \rotatebox{90}{H2H~\cite{h2h_dac}}
  & \rotatebox{90}{\textbf{\sysname{}}}
   \\ \hline
  
  \begin{tabular}{@{}c@{}} Concurrent DNNs\end{tabular} 
  & \multicolumn{1}{c|}\xmark
  & \multicolumn{1}{c|}\xmark 
  & \multicolumn{1}{c|}\xmark 
  & \multicolumn{1}{c|}\cmark
  & \multicolumn{1}{c|}\cmark
  & \multicolumn{1}{c|}\cmark
  & \multicolumn{1}{c|}\cmark
  & \multicolumn{1}{c|}\cmark
  \\ \hline
  
  \begin{tabular}{@{}c@{}} Multi-Accelerator\end{tabular}
  & \multicolumn{1}{c|}\cmark
  & \multicolumn{1}{c|}\cmark
  & \multicolumn{1}{c|}\cmark 
  & \multicolumn{1}{c|}\xmark
  & \multicolumn{1}{c|}\xmark 
  & \multicolumn{1}{c|}\cmark
  & \multicolumn{1}{c|}\cmark
  & \multicolumn{1}{c|}\cmark
  \\ \hline
  
  \begin{tabular}{@{}l@{}} Transition Cost\end{tabular}
  & \multicolumn{1}{c|}\cmark
  & \multicolumn{1}{c|}\cmark
  & \multicolumn{1}{c|}\cmark
  & \multicolumn{1}{c|}\cmark
  & \multicolumn{1}{c|}\cmark
  & \multicolumn{1}{c|}\xmark
  & \multicolumn{1}{c|}\cmark
  & \multicolumn{1}{c|}\cmark
  \\ \hline
  \begin{tabular}{@{}l@{}} Memory Contention\end{tabular}
  & \multicolumn{1}{c|}\xmark
  & \multicolumn{1}{c|}\xmark
  & \multicolumn{1}{c|}\xmark
  & \multicolumn{1}{c|}\xmark
  & \multicolumn{1}{c|}\cmark
  & \multicolumn{1}{c|}\xmark
  & \multicolumn{1}{c|}\xmark
  & \multicolumn{1}{c|}\cmark
  \\ \hline
  \begin{tabular}{@{}l@{}} Dynamic scheduling\end{tabular}
  & \multicolumn{1}{c|}\cmark
  & \multicolumn{1}{c|}\xmark
  & \multicolumn{1}{c|}\cmark 
  & \multicolumn{1}{c|}\xmark
  & \multicolumn{1}{c|}\cmark
  & \multicolumn{1}{c|}\xmark
  & \multicolumn{1}{c|}\xmark
  & \multicolumn{1}{c|}\cmark
  \\ \hline
  \begin{tabular}{@{}l@{}} Optimal schedules \end{tabular}
  & \multicolumn{1}{c|}\xmark
  & \multicolumn{1}{c|}\cmark
  & \multicolumn{1}{c|}\xmark
  & \multicolumn{1}{c|}\cmark
  & \multicolumn{1}{c|}\xmark
  & \multicolumn{1}{c|}\xmark
  & \multicolumn{1}{c|}\xmark
  & \multicolumn{1}{c|}\cmark
  \\ \hline
\end{tabular}
\end{table}
\normalsize

Recently, efficient execution of DNN inference on DSAs has been of interest. 

\textbf{Concurrent-DNN Execution:} Several studies~\cite{kwon2021hpca,h2h_dac,kim2023moca,karatzas2023omniboost, narayanan2020heterogeneity} propose scheduling techniques for the concurrent execution of multiple DNNs. Two of them focus on inference on SoCs: Herald~\cite{kwon2021hpca} introduces a mapper to optimize hardware resource utilization across accelerators such as NVDLA~\cite{nvidia_dla} and Shi-diannao~\cite{du2015shidiannao}. H2H~\cite{h2h_dac} improves on Herald by considering inter-accelerator transition costs. 

\textbf{Multi-accelerator Scheduling:} 
Scheduling for systems with more than one type of accelerators has recently been targeted by a few studies ~\cite{kao2020gamma,kang2020scheduling,wu2020pipeline,boroumand2021google,tzilis2019energy,jeong2021deep,belviranli2016cumas}. 
Among the most relevant, Gamma~\cite{kao2020gamma} and Kang et al.~\cite{kang2020scheduling} build genetic algorithms to utilize multiple accelerators for single DNN execution
while Wu et al.~\cite{wu2020pipeline} and Mensa~\cite{boroumand2021google} targets custom hardware for edge devices. None of these studies address contention and balancing issues with multi-DNN execution. DNN training for large-scale CPU + DSA systems~\cite{narayanan2019pipedream,park2020hetpipe,jiang2020unified,markthub2018dragon}, on the other hand, is outside the scope of this work.



\textbf{Optimal Schedule Generation:} Only a couple of studies, to our knowledge, create optimal schedules for multi-DSA execution. AxoNN~\cite{dagli22AxoNN} maps layers of a single DNN into heterogeneous accelerators under an energy budget, resulting in a serial (\ie{} single DNN) execution. 
OmniBoost~\cite{karatzas2023omniboost} uses Monte Carlo tree search by exhaustively profiling layers on CPU and GPU. Both approaches only create static schedules. 

\textbf{Shared Memory Contention:} None of the studies mentioned so far address shared memory contention. MoCA~\cite{kim2023moca} designs a multitenant DSA architecture 
with dynamic memory resource management. FAST~\cite{zhang2022full} jointly optimizes layer-to-DSA mapping by using ILP-based operator fusion technique to remedy the memory bottlenecks in low compute intensity workloads whereas ParDNN~\cite{qararyah2021computational}  partitions DNNs under a memory limit. However, these approaches are not adaptable to off-the-shelf multi-DSA shared memory SoCs.


Table~\ref{table:related_work} provides a snapshot of what most relevant work offers and how they compare against \sysname{}. Achieving the ideal execution scenario depicted in \textbf{Case 3} of Figure~\ref{fig:proposed_methodology} requires holistic consideration of several factors: (i, ii) Interaction and mapping opportunities created by running \textit{concurrent DNNs} on \textit{different types of DSAs}, (iii) the transition overhead when the execution within a DNN switches across accelerators, (iv) the slowdown caused by the SM contention as layers run concurrently --our analysis shows that SM contention-unaware decisions can reduce system performance by up to 70\%, as detailed in Sec~\ref{sec:multi_dnn}--,
(v) support for dynamic schedules, and (vi) optimal schedule creation. 
The efficient and safe operation of performance critical mobile and autonomous workloads on shared memory SoCs depends on the consideration of all these factors \textit{holistically}. Our experiments demonstrate that the lack of such consideration results in mispredicted performance, which in turn results in inefficient schedules. 


\section{HaX-CoNN: \underline{H}eterogeneity-\underline{a}ware E\underline{x}ecution of \underline{Co}ncurrent Deep \underline{N}eural \underline{N}etworks}

\begin{figure}[t]
\centering
\includegraphics[width=\linewidth]{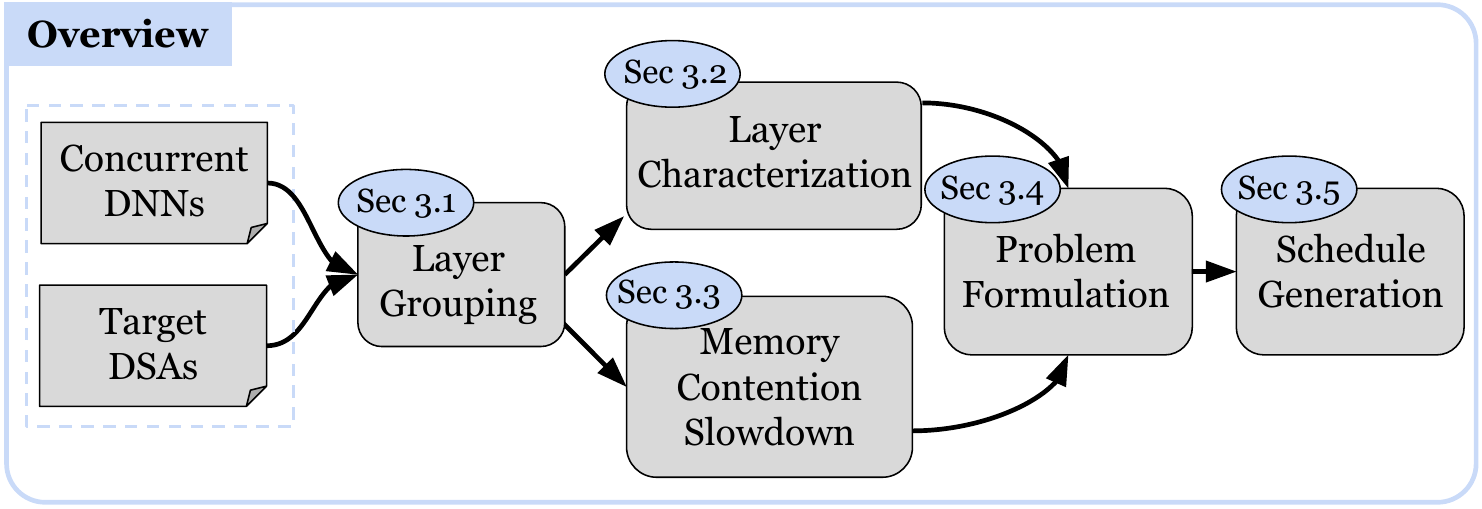}
\vspace{-1.9em}
\caption{\textbf{\sysname{} overview}}
\label{fig:overview}
\vspace{-1.7em}
\end{figure}

An overview of how our proposed methodology works is given in Fig.~\ref{fig:overview}. \sysname{} takes \textit{DNNs} to be scheduled and target \textit{DSAs} as input and produces optimal schedules as output. 

\subsection{Layer Grouping} \label{sec:where_to_split}

The first step involves identifying minimal layer groups to serve as atomic assignment units for DSAs. This grouping considers several factors:

\noindent\textit{1) Preserving layer optimizations:} Layer/operator fusion~\cite{7783725,niu2021dnnfusion,chen2018tvm} merges multiple layers into a single layer, optimizing the computational load and minimizing memory accesses.
\textit{Transition points} during DNN execution, where we switch execution from one DSA to another, should not impede these optimizations. Therefore, we ensure that fusible layers are grouped together and scheduled on the same accelerator.

\noindent\textit{2) Input/output reformatting:} DSAs typically operate in a pipeline. If a transition from a DLA-mapped layer disrupts the internal pipeline, then an additional output reformatting operation is inserted.
Similarly, input reformatting might be required after transitioning from GPU to DLA. Layer groupings can be structured to avoid such formatting overheads.  

\noindent\textit{3) Accelerator and software limitations:}
DSAs are often limited by layer types, layer parameters, batch sizes, etc.
DNN frameworks, such as NVIDIA's TensorRT~\cite{tensorrt} and Qualcomm SNPE~\cite{snpe_qualcomm}, ensure such constraints are followed.  
For example, TensorRT does not allow to perform a transition from DLA to GPU right after the \textit{Eltwise} layer. We identify such limitations via \textit{markOutput} or \textit{canRunOnDLA} TensorRT API calls on NVIDIA. 
Our framework considers the limitations when defining valid transitions between accelerators.

In this step, we group layers as follows:
If transitioning to another DSA after a layer is prohibited or leads to increased overhead, the layer is grouped with subsequent layers. Otherwise, the layer is marked as a potential \textit{transition point}.


\subsection{Per-layer performance and transition characterization} \label{sec:characterization}

After we identify all feasible layer groupings, hence the \textit{transition points}, the next step is to characterize each layer (or group)'s performance and the overhead if an inter-DSA transition occurs after a particular layer.
\vspace{-0.2em}
\paragraph{Layer characterization:}
Strategically assigning layers to the DSAs where they will run most efficiently has the potential to uncover increased performance.     
Previous studies~\cite{kwon2021hpca,hegde2021mind, boroumand2021google, kao2022magma, kao2023optimized} have detailed various parameters that affect the efficiency of deep learning accelerators, such as layer type, input size, kernel size, etc. Different layers within a DNN yield varying performance speedups when run on a specific DSA. To illustrate and analyze this further, we conduct an experiment where we profile layer groups in GoogleNet on GPU and DLA. As shown in Table~\ref{table:characterization}, despite DLA performing slower than GPU for all layers, the speed reduction is less severe for some layers. The fourth column displays the ratio of execution time on DLA over GPU, varying from 1.40x to 2.02x among layer groups (from 1.2x to 3.4x on VGG-19 and from 1.3x to 1.9x on ResNet152). Larger performance discrepancies primarily arise because GPUs, heavily optimized for large-size matrix operations, are capable of more effectively exploiting performance on convolution operations with larger inputs compared to DLA. Conversely, smaller kernels, such as those in groups 95-109 and 124-140, are a better fit for the DLA's internal on-chip buffer.

Prior work~\cite{dagli22AxoNN, karatzas2023omniboost,boroumand2021google,boroumand2021google,jia2022codl,alzubaidi2021review} show that it is feasible to characterize DNNs via a \textit{layer-centric} profiling approach where commonly used layer types are profiled beforehand for different input and filter sizes.  Following a similar methodology, we profile each layer or layer group on the DSAs in the system. 
We utilize IProfiler interface of TensorRT on NVIDIA devices~\cite{tensorrt_iprofiler}, which reports per layer time.
Profiled execution times are later then embedded into variable $t$, which is used in the equations ~\ref{eq:time},~\ref{eq:start_time}, ~\ref{eq:end_time}, and ~\ref{eq:slowdown} in Sec~\ref{sec:formulation}. 
 Overall, the number of layer/input combinations we use in our profiling is 225 per accelerator. It is essential to note that this offline profiling step is done only once for each accelerator on the target platform. Since our approach is layer-centric (\ie{} DNN independent), it is not necessary to repeat the profiling process for a specific DNN, regardless of whether it has been encountered before or not.

\vspace{-0.4em}
\paragraph{Inter-DSA layer transitions:} 
Despite the potential performance boost offered by multi-DSA execution, transitioning between DSAs comes with a cost. This cost, crucial for accurate performance predictions and optimal scheduling, is contingent on the
size of the transient data in private caches of DSAs. The output of the layer preceding the transition is flushed back to the shared memory so that the DSA where the next layer will execute on can access it.  
The fifth column in Table \ref{table:characterization} represents the time spent when the execution switches from GPU to DLA after each layer group. As output data sizes decrease toward the end of layer groups, so does transition time. 
Notably, our experiments also reveal that some layer groups, such as 39-53 and 95-109, which end with pooling layers result in significantly less transition overhead.
We empirically derive the transition costs of the layers on our target set of accelerators, following the methodology outlined in ~\cite{dagli22AxoNN}. To implement them, we insert $MarkOutput$ and $addInput$ API calls in TensorRT~\cite{tensorrt_iprofiler}. We then incorporate them into Eq. ~\ref{eq:time} and ~\ref{eq:transition} in Sec~\ref{sec:formulation}.


\normalsize

\begin{table}[t]
\footnotesize
\centering
\caption{Execution and transition time of layer groups in GoogleNet}
\vspace{-0.3cm}
\label{table:characterization}
\begin{tabular}{ | m{1.1cm} | c| c | c| c |c|  }
  \hline
  Layer Group & \begin{tabular}{@{}c@{}}GPU \\ (ms) \end{tabular}  & \begin{tabular}{@{}c@{}}DLA \\ (ms) \end{tabular}   & \begin{tabular}{@{}c@{}}\footnotesize{D/G Exec} \\ \footnotesize{Time} \\ \footnotesize{Ratio}  \end{tabular} & \begin{tabular}{@{}c@{}}\footnotesize{Transition} \\ \footnotesize{Time From} \\ \footnotesize{ G to D (ms)}  \end{tabular}  & \begin{tabular}{@{}c@{}}Memory \\ Thr. (\%) \end{tabular}   \\ \hline
0-9     &   0.45    & 0.75 &    1.65 & 0.056 &  41.97 \\ \hline
10-24   &   0.19    & 0.34 & 	1.80 & 0.075 & 	62.21 \\ \hline
25-38   & 	0.31    & 0.45 & 	1.44 & 0.062 & 	78.49 \\ \hline
39-53   & 	0.18	& 0.37 & 	2.02 & 0.011 & 	53.41 \\ \hline
52-66   & 	0.16	& 0.31 & 	1.98 & 0.055 & 	55.70 \\ \hline
67-80   & 	0.17	& 0.33 & 	1.96 & 0.024 & 	59.24 \\ \hline
81-94   & 	0.21	& 0.31 & 	1.50 & 0.058 & 	62.60 \\ \hline
95-109  & 	0.25	& 0.35 & 	1.40 & 0.03  & 	76.12 \\ \hline
110-123 & 	0.16	& 0.27 & 	1.66 & 0.024 & 	66.95 \\ \hline
124-140 & 	0.24	& 0.36 & 	1.49 & 0.007 & 	47.96 \\ \hline
\end{tabular}
\vspace{-2em}
\end{table}
\normalsize

\subsection{Characterizing shared-memory contention}
\label{sec:contention}



One of the core novelties of our works is its ability to account for the slowdown caused by shared memory contention. Since existing multi-DSA schedulers do not consider this when making scheduling decisions, the resulting mappings often leave the system under-utilized. However, estimating this slowdown, especially for multi-DSA systems, is not trivial. If we were to build upon a layer-wise profiling approach, we would need to perform exhaustive and peer-wise runs of all combinations of layers that can be collocated. This will result in a factorial explosion of profiling search space and require significant profiling time~\cite{zhu2017co}. 

To prevent this, we follow a \textit{decoupled} two-step approach: we first characterize each layer's requested memory throughput when they are run standalone. Using these throughput values, we then utilize a processor-centric slowdown model, PCCS~\cite{Pccs_micro}, to estimate the slowdown without relying on layer-specific information. PCCS represents the slowdown between multiple concurrent workloads as a function of requested memory throughput and external memory traffic, and builds a piece-wise model to predict the slowdown experienced by the accelerator requesting the throughput. 
Built upon PCCS, our decoupled approach is performed at layer-level and separates the collection of layer-specific standalone performance profiles, as collected in Sec~\ref{sec:characterization}, and the slowdown caused by concurrent execution. 
The last column in Table~\ref{table:characterization} lists memory throughput measurements per layer group in GoogleNet. The general pattern we observe in many DNNs is that  higher input size results in higher memory throughput. We also observe that, as the filter size in convolution and pooling layers gets larger, there is a decrease in throughput due to the increasing arithmetic intensity of the underlying operation.

\begin{figure}
\vspace{-0.5em}
\centering

\includegraphics[width=1\linewidth]{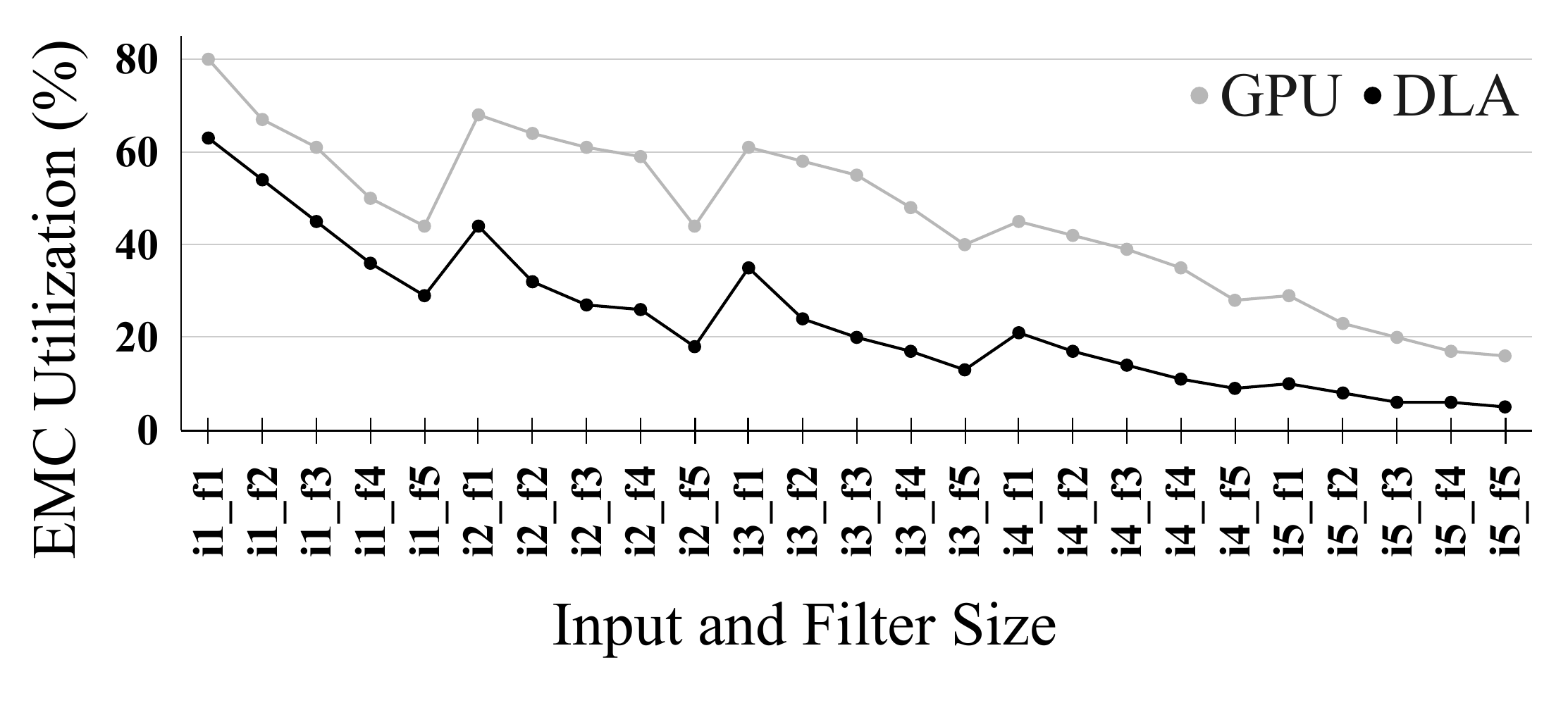}
\vspace{-2.5em}
\caption{EMC utilization by conv layers on GPU and DLA}
\vspace{-1.7em}
\label{fig:emc_gpu_dla}
\end{figure}


Conventional hardware counters to monitor requested memory throughput may not be applicable for some \textit{black-box} DSAs which cannot be profiled with conventional tools -- NVIDIA’s Nsight Compute tool~\cite{nsight_compute} can profile requested memory throughput on GPUs but not on DLAs. As an alternative way to methodologically solve this issue, we develop a four-step approach: 1) We first profile target layers on GPU and analyze the memory throughput for several layer types (\ie{} convolution, pooling, and fully connected) and their parameters (\ie{} input size filter size). Throughout their execution lifetimes, we observe that many layers individually exhibit homogeneous memory access characteristics as they internally embed homogeneous and dense computations. 2) We then profile external memory controller (EMC) utilization for all layers on both DLA and GPU. Our analysis, which is shown in Fig~\ref{fig:emc_gpu_dla}, reveals that the EMC utilization curves for DLA and GPU are correlated and proportional. 3) Using this observation, for a given layer, we estimate its memory throughput on black-box DSAs (\eg{} DLAs) by dividing its GPU-based throughput by the ratio of EMC utilization of GPU and DSA for that specific layer. 4) Finally, by utilizing PCCS, we estimate the slowdown of a layer via its requested memory throughput and the external memory throughput requested by the other concurrently running layer on the other DSA. 


\begin{figure}[t]
\centering

\includegraphics[width=0.7\linewidth]{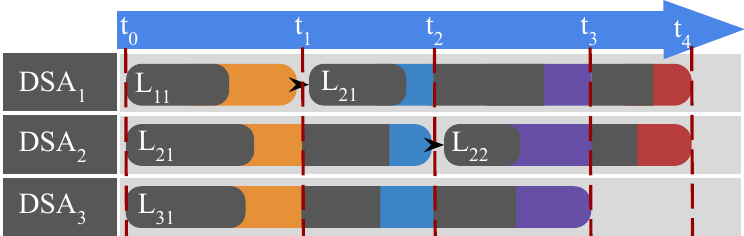}
\vspace{-0.4em}
\caption{\textbf{Illustration for a hypothetical execution of five layers from three DNNs running on three different accelerators. Colored regions indicate additional slowdowns that each layer experiences for varying amounts of external memory pressure.  
}}
\vspace{-1.5em}
\label{fig:cfg_contention}
\end{figure}

When multiple layers are run simultaneously on different accelerators, the degree of slowdown throughout their execution is non-uniform and depends on the other layers running concurrently. Fig. \ref{fig:cfg_contention} illustrates this behavior by depicting the execution timelines of five hypothetical layers belonging to three different DNNs. Each timeline represents the execution of $i$th layer on $j$th DNN, labeled as $L_{ij}$ on $DSA_k$. The black regions in the timeline represent the time that the executions would take if all layers were run separately. Each colored extension to the black regions indicates slowdowns for different sets of layers running together. To address the complexity of handling varying amounts of slowdown during the execution of collocated layers, we introduce a scheduling concept called \textit{contention interval}. Each contention interval ($t_i$, $t_{i+1}$), represents a period during a layer starts or ends, which is embedded into Eq.~\ref{eq:interval} in Sec~\ref{sec:formulation}. During each contention interval, different rates of slowdowns are observed by each layer, and the slowdown depends on the cumulative external memory pressure demands during that interval.

\subsection{Formulating the problem}\label{sec:formulation}
We integrate layer execution time, inter-accelerator transition time, and memory contention slowdown into an empirical model and formulate the scheduling problem as a series of linear equations. 
Table~\ref{table:notation_1} summarizes the variables and notations we use in the formulation. 
The primary input to our model is the $DNN$ set for which we explore its mapping to the accelerator set $A$. $L_{i,n}$ denotes the smallest schedulable layer entity that belongs to the layer set of $DNN_{n}$. A layer entity is either a single layer or a group of layers, as identified by Section~\ref{sec:where_to_split}. Functions  $t(L_{i,n},a)$ and $\tau(L_{i,n},a,OUT|IN)$ represent the execution time and transition overheads of layer $L_{i,n}$ on accelerator $A_a$.

The goal of our formulation is to find a schedule $S$ for all layers across all DNNs. The schedule function, defined in Eq.~\ref{eq:scheduling_1}, returns $A_a$ that $L_{i,n}$ is mapped to.  $S$ is assumed to be initially unknown and will be determined  by the solver later.

\vspace{-0.6em}
\begin{equation}
\begin{aligned}
S(L_{i,n}) = A_a  \; \; where \;\;\; 1\leq i \leq len(DNN_n) ,  \\
 \; \; 1\leq n \leq len(DNN) \; \; ,  \; \; 1\leq a \leq len(A)
\end{aligned}
\label{eq:scheduling_1}
\vspace{-0.3em}
\end{equation}

Total execution time of a DNN is formulated in Eq.~\ref{eq:time}. Total time comprises \textit{standalone execution time} $t$ of each layer, the \textit{slowdown} $C$, and and $IN$ and $OUT$ \textit{transition costs}, $\tau$. 
\vspace{-0.5em}
\begin{equation}
\begin{aligned}
T(L, S(L \rightarrow A))_{n} =\sum_{i=0}^{len(DNN_{n})} t(L_{i,n},S(L_{i,n})) *C_{L_{i,n},S(L),L}  \\ +  \; \;  TR_{i,n} \; \times \; \tau(L_i, s(L_i), OUT)  + \; \; TR_{i,n} \; \times \;  \tau(L_{i+1},s(L_{i+1}),IN)
\label{eq:time}
\end{aligned}
\end{equation}

\begin{table}[t]
\caption{The notation used by the formulation.}
\vspace{-1em}
\centering
\begin{center}

\footnotesize\begin{tabular}{|p{1.5cm}|p{6cm}|}

\hline
    \textbf{Notation} & \textbf{Explanation}  \\ \hline 
$DNN_{n}$ & $n$th DNN in the given $DNN$ set which contains networks to be executed concurrently \\ \hline
$L_{i,n}$ & $i$th layer of the $n$th DNN in the $DNN$ set\\ \hline
$len(DNN_{n})$ & total number (length) of layer groups in $DNN_{n}$ \\ \hline
$A_a$ & $a$th accelerator in the given accelerator set $A$
\\ \hline
S($L_{i,n}$) &  The schedule, \ie{} accelerator mapping, of $L_{i,n}$ \\ \hline
$t(L_{i,n},A_a)$ &  Total execution time of $L_{i,n}$ on $A_a$  \\ \hline
$st(i,n)$ & Execution start time of $L_{i,n}$ \\ \hline
$et(i,n)$ & Execution end time of $L_{i,n}$ \\ \hline

\begin{tabular}{@{}l@{}}$\tau(L_{i,n},A_a, $ \\ $OUT|IN)$ \end{tabular}  & \begin{tabular}{@{}l@{}}The time required to transition the DNN execution \\ after$|$before layer $l_i$ executed on accelerator $A_a$ \end{tabular}     \\ \hline 
$TR_{i,n}$ & Boolean var if a transition is set after layer $L_{i,n}$ \\ \hline
$T(L,S(L))_{n}$& Total execution time elapsed by the execution of given sets of layer $L$ of the $n$th DNN \\ \hline

$c_{L_{i,n},S(L),L}$ & The slowdown of $L_{i,n}$ due to the contention caused layers running on other accelerators, \ie{} $S(L)$ \\ \hline
$I_{i,j}$ & The length of interval where layers $i$ and $j$ overlap  \\ \hline
$Int$ & Interval array holding start and end time  of layers \\ \hline
\end{tabular}
\vspace{-2em}
\end{center}
\label{table:notation_1}
\end{table}
\normalsize

We encode the decision to make transitions into our formulation via the boolean function given in Eq.~\ref{eq:transition}.
This function compares the accelerator assignments of adjacent layers $L_{i,n}$ and $L_{i+1,n}$. If the assignments differ, the transition cost, $\tau$, is subsequently incorporated into Eq.~\ref{eq:time}.
\vspace{-0.5em}
\begin{equation}
\begin{aligned} 
TR_{i,n}=
\begin{cases}
			\text{1 \;\;\;\;\;, if \; $S(L_{i,n})$ $\neq$ $S(L_{i+1,n})$} \;\; \\
			\text{0 \;\;\;\;\;, if \; $S(L_{i,n})$ = $S(L_{i+1,n})$}
		 \end{cases}
\label{eq:transition}
\vspace{-0.4em}
\end{aligned} 
\end{equation}


Eq.~\ref{eq:start_time}~and~\ref{eq:end_time} computes the execution start and end times, $st()$ and $et()$ respectively, for layer $L_{i,n}$. $Int$ array in Eq.~\ref{eq:interval_set} stores the start and end time for layers, facilitating the iterative comparison of the contention intervals across layers.  

\vspace{-1em}
\begin{equation}
st(i,n) = T(L_{0\; to\; i-1 \;,\; n}, S(L))_{n} 
\label{eq:start_time}
\vspace{-1em}
\end{equation}

\begin{equation}
et(i,n) = st(i,n) + t(L_{i,n},S(L_{i,n})) * C_{i,n}
\label{eq:end_time}
\end{equation}

\vspace{-1em}
\begin{equation}
\begin{aligned}
\forall L_{i,n} \;\;, \;\; [st(i,n),et(i,n)] \in Int \quad \quad \quad \quad  \\
where \;\;\; 1\leq i \leq len(DNN_{n}) ,  \; \; 1\leq n \leq len(DNN) \; \;
\label{eq:interval_set}
\end{aligned}
\vspace{-0.5em}
\end{equation}

The contention function $C$, outlined in Eq.~\ref{eq:slowdown}, calculates the total slowdown for layer $L_{i,n}$ by taking each CI time overlapping with that layer and the slowdown ratio corresponding to the interval. The contention model, $cont\_model$ returns an estimated slowdown amount depending on bandwidth demand by layer $l_i$ and cumulative external bandwidth demand by other layers running on the same interval. 
            


\vspace{-1em}
\begin{equation}
\begin{aligned} 
C_{L_{i,n},S(L),L}= \sum_{I_k \in Int}^{} \frac{I(L_{i,n}\;,\; L_{j,n}) * cont\_model(L_{i,n}\;,\; L_s)} {t(L_{i,n},S(L_{i,n})) * len(L_s)} 
\; \\ 
where \; 1\leq j \leq len(DNN_n) ,  \; 1\leq n \leq len(DNN) ,\; L_{j,n} \; \in \; L_s \\ Int_k\; \cap \; [st_{i,n},et_{i,n}] \; \neq \; \emptyset \;,\; Int_k \cap [st_{j,n},et_{j,n}] \neq \; \emptyset 
\label{eq:slowdown}
\end{aligned}
\end{equation}

Eq.~\ref{eq:interval} details how we determine the start and end of a CI, based on the start and end time of concurrently running layers. If a layer faces no contention, the equation simply returns the layer's execution time, leading to a value of 1 to be returned in Eq.~\ref{eq:slowdown}, thereby indicating no slowdown effect for a layer running independently in Eq.~\ref{eq:time}.
\vspace{-0.5em}
\begin{equation}
\begin{aligned} 
I(i,j)=\begin{cases}

            e_j-s_i & \;\; if (s_j \leq s_i \leq e_j \;\;\; \& \;\;\; s_i \leq s_j \leq e_i)  \\
            
            e_j-s_j & \;\; if (s_i \leq s_j \leq e_i \;\;\; \& \;\;\; s_i \leq s_j \leq e_i)  \\
            
            e_i-s_j & \;\; if (s_i \leq s_j \leq e_i \;\;\; \& \;\;\; s_j \leq e_i \leq e_j) \\
              
            e_i-s_i & \;\; if (s_i \leq s_j \;\;\; \& \;\;\; e_i \leq e_j)\\
            
            e_i-s_i & \;\;  otherwise
		 \end{cases}
\end{aligned}
\label{eq:interval}
\vspace{-0.1em}
\end{equation}
            
            
              

We establish a constraint in Eq.~\ref{eq:no_overlap} that limits two distinct layers from sharing the same accelerator for longer than an $\varepsilon$ interval. 
Ideally, in a flawless model, the estimated execution and slowdown values could yield perfect transitions where accelerator usage periods can be precisely predicted. Variable $\varepsilon$ allows us to mitigate the prediction error,  
and facilitates more transition points by allowing for a tiny overlap of concurrently assigned layers on the same accelerator at the start or end of their executions. 
\begin{equation}
\begin{aligned}
\nexists \; L_{i,nn},L_{j,n} \;(L_{i,nn} \in DNN_{nn} \;\; and \;\; L_{j,n} \in DNN_n \;\;|\quad\\ st_{L_{j,n}} \!<\! st_{L_{i,nn}}\! \mp \varepsilon < et_{L_{i,nn}} \; or \; st_{L_{j,n}}\! < et_{L_{i,nn}} \mp \varepsilon < et_{L_{j,n}} ) \\ where \; S(L_{i,nn})= S(L_{j,n}) \;\;, \;\; nn\neq n\;\;\;\; \quad\quad\quad 
\label{eq:no_overlap}
\end{aligned}
\end{equation}





\paragraph{Objective functions: } Depending on the different scenarios that a user may target, we propose two separate objective functions: Equation~\ref{eq:max_thr} maximizes the utilization of the system to increase the total throughput and Equation~\ref{eq:minmax} minimizes the maximum latency among DNNs. 
The use cases for objective functions are further elaborated via scenarios in Section~\ref{sec:scenarios}.

\vspace{-2em}
\setlength{\columnsep}{-0.2cm}
\begin{multicols}{2}
\begin{equation}
\vspace{-1em}
\!\!\!\!\!\!\max \!\!\!\! \sum_{n=1}^{len(DNN)} \!\!\!\! \frac{1}{T(L,S(L))_{n}} \!\!\!\!\!\!\!\!\!\!
\label{eq:max_thr} \\
\end{equation}\break
\begin{equation}
\vspace{-0.5em}
\min \max  \; T(L,S(L))_{n}\!\!\!
\label{eq:minmax} \\
\end{equation}
\vspace{-3.5em}
\end{multicols}

\subsection{Optimal and Dynamic Schedule Generation}\label{sec:objective}
In our work, we only produce optimal schedules that satisfy given objectives and constraints because we don’t resort to heuristics to find such schedules. We achieve this by representing the entire scheduling problem formulated in Section \ref{sec:formulation} as a constraint-based optimization problem and solving with industry-strength SAT solvers that embeds decades of engineering to quickly find optimal solutions. 
Advanced SMT solvers, such as Z3~\cite{de2008z3}, Gurobi~\cite{gurobi}, and OptiMathSAT~\cite{sebastiani2015optimathsat}, employ branch \& bound techniques to converge towards optimal solutions for many NP-complete problems (\ie{} job-shop scheduling)~\cite{roselli2018smt}. Considering the relatively small parameter search space of our targeted problem set(\ie{} total number of accelerators and tasks in the system), the use of SMT solvers provides optimal schedules in seconds. Depending on the operational requirements of the autonomous system, optimal schedules can be explored either statically or dynamically. 

Generating optimal schedules beforehand (\ie{} \textit{statically}) is feasible for a variety of scenarios, such as in autonomous systems with fixed resolution input devices (like cameras and lidars) and many DNNs designed for fixed image or video frame sizes. Some scenarios, such as a drone switching between \textit{discovery} or \textit{tracking} modes, might require unique CFGs. Such CFGs (or the path followed in a CFG) and their corresponding schedules can be pre-determined statically and toggled during the execution. Thus, users of \sysname{} can rely on offline profiling to determine the execution costs needed for static scheduling~\cite{zou2022resilience}. 

There are other cases where the static generation of optimal schedules may not be possible. For example, different DNN models may be required for various phases of the autonomous system execution~\cite{boroujerdian2021roborun,zhao2020safety,hsiao2022zhuyi}, resulting in an unpredictable change in the CFG. For such scenarios, we propose D-\sysname{}, a runtime-based adaptation of our solution to (1) run SAT solvers on-the-fly, (2) gradually achieve and apply better schedules, and (3) eventually reach an optimal solution as the autonomous system continues to operate. This approach is feasible because autonomous systems often embed long-running loops and once an optimal schedule is found for a newly changed CFG, it will be reused for a while.

\begin{table}[b]
\vspace{-1em}
\footnotesize
\caption{\textbf{The HW specifications for targeted architectures.}}
\label{table:socs}
\vspace{-0.8em}
\begin{tabular}{|m{1.4cm}|m{6cm}|}
\hline
\multicolumn{2}{|c|}{ \textbf{NVIDIA AGX Orin}} 
\\ \hline
GPU  & Ampere arch. 1792 CUDA \& 64 Tensor cores   \\ \hline
DSA & NVDLA v2.0 \\ \hline
CPU   & 12-core Arm Cortex v8.2 64-bit  \\ \hline
Memory   & 32GB  LPDDR5  | \textbf{Bandwidth: 204.8 GB/s} with 256-bit  \\ \hline
Software & JetPack 5.0.1  \\ \hline
\end{tabular}
\begin{tabular}{|m{1.4cm}|m{6cm}|}
\hline
\multicolumn{2}{|c|}{\textbf{NVIDIA Xavier AGX} } 
\\ \hline
GPU  &     Volta arch. 512 CUDA and 64 Tensor cores    \\ \hline
DSA & NVDLA v1.0 \\ \hline
CPU & 8-core Carmel Arm v8.2 64-bit \\ \hline
Memory &  16GB LPDDR4 | \textbf{Bandwidth: 136.5 GB/s}, 256-bit \\ \hline
Software & JetPack 4.5  \\ \hline
\end{tabular}
\begin{tabular}{|m{1.4cm}|m{6cm}|}
\hline
\multicolumn{2}{|c|}{\textbf{Qualcomm 865 Mobile Development Kit} } 
\\ \hline
GPU  &  Qualcomm Adreno™ 650 GPU    \\ \hline
DSA & Hexagon 698 DSP \\ \hline
CPU & Qualcomm  Kryo 585, 8-core, up to 2.84GHz \\ \hline
Memory & 6GB LPDDR5 | \textbf{Bandwidth: 34.1 GB/s} with 64 bits \\ \hline
\end{tabular}
\end{table}

\section{Experimental Setup}
\label{sec::experimental_setup}

\paragraph{Computing platforms:} We use three popular heterogeneous SoCs to evaluate \sysname{}: NVIDIA's AGX Orin~\cite{orin_bib}, Xavier AGX~\cite{xavier_agx_bib}, and Qualcomm 865 Development kit~\cite{snapdragon_865}. All three platforms have a shared memory system with multiple accelerators. The technical specifications of these systems are summarized in Table \ref{table:socs}. 
It is essential to note that the maximum number of accelerators we consider in our experiments is limited to two, because, to the best of our knowledge, there are no off-the-shelf SoCs that offer more than two types of programmable DSAs for DNN acceleration. 

\begin{table}[t]
\small
\caption{\textbf{Standalone runtimes (ms) and relative performances }}
\vspace{-0.8em}
\begin{tabular}{|m{1.5cm}|m{1cm}m{1cm}|m{1cm}m{1cm}|}
\hline
\diagbox[width=5.8em]{DNN}{ Device} & \multicolumn{2}{c}{\textbf{NVIDIA AGX Orin}} & \multicolumn{2}{|c|}{\textbf{NVIDIA Xavier AGX}}
\\ \hline 

& \begin{tabular}{@{}l@{}}GPU \\ (ms) \end{tabular}  
& \begin{tabular}{@{}l@{}}DLA \\ (ms) \end{tabular}
& \begin{tabular}{@{}l@{}}GPU \\ (ms) \end{tabular}  
& \begin{tabular}{@{}l@{}}DLA \\ (ms) \end{tabular} 
\\ \hline

\textbf{CaffeNet}    &0.74  &1.79 &2.26 &5.51    \\ \hline
\textbf{DenseNet}    &2.19  &3.10 &7.84 &-    \\ \hline
\textbf{GoogleNet}   &0.99 & 1.52 & 1.98 & 3.68   \\ \hline
\textbf{Inc-res-v2}  &3.06  &5.15 &15.12 &17.95    \\ \hline
\textbf{Inception}   &2.49  &5.66 &8.31 &15.94    \\ \hline
\textbf{ResNet18}    &0.41  &0.74 &1.37 &2.81    \\ \hline
\textbf{ResNet50}    &0.91  &1.67 &2.88 &6.01    \\ \hline
\textbf{ResNet101}   &1.56  &2.47 &5.34 &10.6    \\ \hline
\textbf{ResNet152}   &2.19  &3.26 &7.7 &12.71    \\ \hline
\textbf{VGG19}       &1.07  &2.93 &5.95 &19.05    \\ \hline

\end{tabular}
\vspace{-1em}
\label{table:standalone}
\end{table}

\paragraph{Applications:} We use the DNNs that are commonly used in benchmarking DNN inference:  Alexnet ~\cite{krizhevsky2012imagenet}, GoogleNet\cite{szegedy2015going}, Inception-V4\cite{szegedy2017inception},  ResNet18/52/101/152~\cite{He_2016_CVPR}, VGG-19~\cite{simonyan2015vgg}, FCN-ResNet18, CaffeNet~\cite{jia2014caffe}, DenseNet~\cite{huang2017densely}, and Inc-Res-v2~\cite{Szegedy2017Inceptionv4IA} with datasets from COCO~\cite{lin2014microsoft}, ImageNet ILSVRC~\cite{russakovsky2015imagenet}, and Cityshape~\cite{cordts2016cityscapes}. These DNNs could be used for various tasks in autonomous systems, such as object detection, image recognition, semantic segmentation, pose estimation, and depth estimation~\cite{grigorescu2020survey}. 

\paragraph{Profiling:} Profiling duration varies by platforms: Computation, transition, and contention characterizations can take up to 3, 10, and 15 minutes, respectively, on NVIDIA Orin, Xavier, and Qualcomm platforms. Since our approach is layer-centric, we performed profiling only once and it is offline.
 

\paragraph{Neural network synchronization:} TensorRT natively does not provide support synchronization between the layers of DNNs concurrently running at different DSAs.
To make sure that the inter-accelerator transitions across DNNs are properly performed, we implement a TensorRT plugin that employs inter-DNN synchronization via inter-process shared memory primitives.


\paragraph{Schedule generation:}  We solve our formulation given in Section \ref{sec:formulation} by using Z3 SMT solver. Z3 has an API in Python~\cite{z3_python} and has shown superior performance for scheduling problems over popular solvers~\cite{roselli2018smt}. Z3 works by determining the satisfiability of the constraints and finding an optimal solution to a given objective. In almost all experiments, Z3 takes under three seconds to run on a single CPU core of NVIDIA Orin AGX. In some rare cases, such as for the Inception-ResNet-v2 network, which consists of 985 layers, the solver takes around ten seconds to find a schedule.

\section{Evaluation}
\label{sec:scenarios}
We demonstrate the utility of \sysname{} via four execution scenarios with different objectives and also via an experiment that exhaustively collocates all the DNNs in our evaluation set.  
Scenario 1 aims to maximize throughput in concurrent data processing on the same DNN whereas scenarios 2 and 3 target two different DNNs operating in parallel and in a pipeline fashion, respectively. Scenario 4 is a hybrid of scenarios 2 and 3. We benchmark \sysname{} against five different baselines: (1) GPU only, (2) non-collaborative GPU \& DLA, (3) Mensa~\cite{boroumand2021google} (which supports single DNN execution only), (4) Herald~\cite{kwon2021hpca}, and (5) H2H~\cite{h2h_dac} (which both support multi-DNN execution). 

\subsection{Running Multiple Instances of the Same DNN }\label{sec:single_dnn}

\begin{figure}[b]
\includegraphics[width=0.95\linewidth]{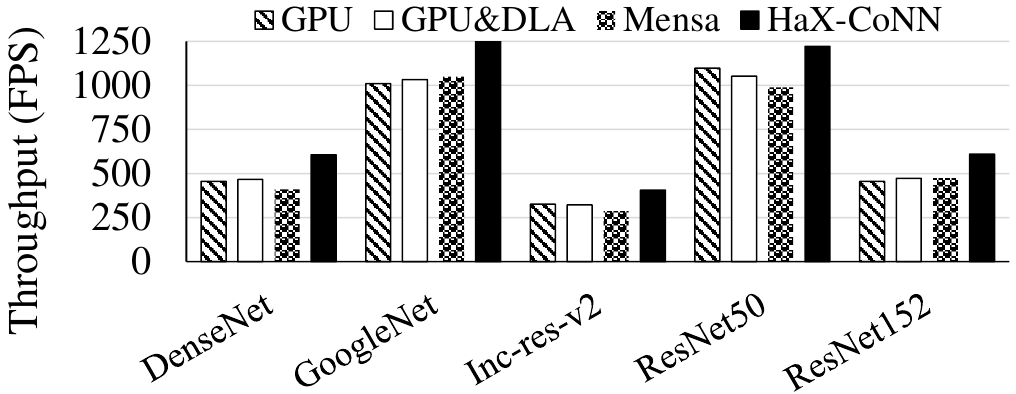}\\
\vspace{-0.6em}
\caption{Throughput (FPS) comparison for Scenario 1: Multiple instances of
the same DNN is run concurrently on NVIDIA AGX Orin}
\label{fig:single_nn_results}
\end{figure}

\begin{table*}[t]
\centering
\caption{Experiment designs for Scenarios 2, 3, and 4, and comparison against baselines on NVIDIA Xavier AGX in experiments 1-5, NVIDIA AGX Orin in experiments 6-8, and Qualcomm 865 in experiments 9-10. DSA refers to DLA for NVIDIA platforms and to the Hexagon DSP for Qualcomm platform. }
\footnotesize
\vspace{-0.3cm}
\begin{tabular}{|m{1.5em}|m{3.3em}|m{4.3em}|m{4.3em}|m{1.6em}m{1.6em}|m{1.6em}m{1.6em}|m{1.6em}m{1.6em}|m{1.6em}m{1.6em}|m{1.4em}m{1.8em}|m{2.2em}m{1.9em}|m{1.9em}m{1.7em}|}\hline
\multirow{2}{*}{\begin{tabular}{@{}c@{}} Exp \\ \#   \end{tabular}}& \multirow{2}{*}{Goal}     & \multirow{2}{*}{DNN-1}   & \multirow{2}{*}{DNN-2}   
&\multicolumn{2}{c|}{\begin{tabular}{@{}c@{}} (1) GPU \\ only \end{tabular}}  &\multicolumn{2}{c|}{\begin{tabular}{@{}c@{}} (2) GPU \& \\ DSA   \end{tabular}}
&\multicolumn{2}{c|}{(3) Herald} 
&\multicolumn{2}{c|}{(4) H2H} &\multicolumn{2}{c|}{\begin{tabular}{@{}c@{}}Optimal \\ schedule  by \\ HaX-CoNN \end{tabular}}
&\multicolumn{2}{c|}{\begin{tabular}{@{}c@{}} Runtime of \\ \sysname{}\\ schedule   \end{tabular}}
&\multicolumn{2}{c|}{\begin{tabular}{@{}c@{}}Improvement \\ over the best \\ baseline (\%) \end{tabular}} \\ \cline{5-18}
&&&&Lat. &FPS &Lat. &FPS &Lat. &FPS &Lat. &FPS  &TR &Dir.  &Lat. &FPS &Lat. &FPS   \\\cline{1-18}
1
 
& \begin{tabular}{@{}l@{}} Min \\ Latency   \end{tabular}   & VGG-19 & ResNet152   
&17.05 
&58 
&\textbf{16.05} 
&62 
& 19.73
& 50
& 16.55
& 60
&\scriptsize{\begin{tabular}{@{}l@{}}29 \\ 89\end{tabular}} 
& \scriptsize{\begin{tabular}{@{}l@{}}DtoG \\ GtoD\end{tabular}} 
&\textbf{13.01} 
& 77
& 23
& 22 \\ \hline
2 
& \begin{tabular}{@{}l@{}} Min \\ Latency   \end{tabular}  & ResNet152 & Inception  
&16.23 
&61  
&15.96 
&62 
& 15.81
& 63
& \textbf{15.75}
& 64  
&\scriptsize{\begin{tabular}{@{}l@{}}188 \\ 72\end{tabular}} 
& \scriptsize{\begin{tabular}{@{}l@{}}DtoG \\ GtoD\end{tabular}}   
& \textbf{13.11} 
& 76
& 20
& 18 \\ \hline

3
& \begin{tabular}{@{}l@{}} Max \\ FPS   \end{tabular}  & Alexnet   & ResNet101 
&11.04 
&90  
&10.97 
&\textbf{93} 
& 12.10
& 82
& 11.49
& 87
&\scriptsize{\begin{tabular}{@{}l@{}}11 \\ 161\end{tabular}} 
& \scriptsize{\begin{tabular}{@{}l@{}}GtoD \\ DtoG\end{tabular}}   
& 8.7
& \textbf{115}
& 26
& 23 \\ \hline

4  %
& \begin{tabular}{@{}l@{}} Max \\ FPS   \end{tabular} & ResNet101 & GoogleNet  
&7.02 
&\textbf{143} 
&7.37 
&140 
&8.95
&111
&9.10
&109
&\scriptsize{\begin{tabular}{@{}c@{}}0 \\ 0\end{tabular}} 
& \scriptsize{\begin{tabular}{@{}c@{}}DtoG \\ DtoG\end{tabular}}   
&7.02  
&\textbf{143}
&0
&0 \\ \hline
5& \begin{tabular}{@{}l@{}} Min \\ Latency   \end{tabular}  & \footnotesize{\begin{tabular}{@{}c@{}}GoogleNet \\ ResNet152\end{tabular}}    & FC\_ResN18    %
&\textbf{15.41} 
&77  
&18.88 
&61 
& 23.68
& 47
& 20.90
& 54
&\scriptsize{\begin{tabular}{@{}l@{}}38 \\ 235\end{tabular}} 
& \scriptsize{\begin{tabular}{@{}l@{}}DtoG \\ GtoD\end{tabular}} 
&\textbf{12.09}  %
&85
&22
&21 \\ \hline
6
& \begin{tabular}{@{}l@{}} Min \\ Latency   \end{tabular}  & VGG-19 & ResNet152        
&\textbf{3.95} 
& 267 
& 4.58
& 218
& 5.76  
& 174  
& 4.90 
& 204 
&\scriptsize{\begin{tabular}{@{}l@{}}27 \\ 95\end{tabular}} 
& \scriptsize{\begin{tabular}{@{}l@{}}DtoG \\ GtoD\end{tabular}}   
&\textbf{3.21}
& 311
& 23
& 22\\ \hline
7  & \begin{tabular}{@{}l@{}} Max \\ FPS   \end{tabular} & GoogleNet& ResNet101  
&4.12 
&378  
&4.24 
&364   
& 4.44
& 340
& 4.13
& \textbf{380}
&\scriptsize{\begin{tabular}{@{}l@{}}38 \\128 \end{tabular}} 
& \scriptsize{\begin{tabular}{@{}l@{}}DtoG \\ GtoD\end{tabular}}   
& 3.4 
& \textbf{426}
& 19
& 18 \\ \hline
8 & \begin{tabular}{@{}l@{}} Min \\ Latency   \end{tabular}  & \footnotesize{\begin{tabular}{@{}c@{}} ResNet101\\GoogleNet\end{tabular}}    & Inception    %
& 5.06 
& 197
& 4.97
& 201
& 5.56
& 180
& \textbf{4.91}
& 203
&\scriptsize{\begin{tabular}{@{}l@{}}31 \\ 88\end{tabular}} 
& \scriptsize{\begin{tabular}{@{}l@{}}DtoG \\ GtoD\end{tabular}}
& \textbf{4.41}
& 226
& 13
& 12 \\ \hline
9 & \begin{tabular}{@{}l@{}} Max \\ FPS   \end{tabular} & GoogleNet &ResNet101 & 98.3 &10.1 &79.1 &\textbf{12.6} &95.9&10.4 &113.8 &8.8 &\scriptsize{\begin{tabular}{@{}l@{}}52 \\ 148\end{tabular}} & \scriptsize{\begin{tabular}{@{}l@{}}DtoG \\ GtoD\end{tabular}} 
& 71.08 &  \textbf{14.1}
& 11 & 10 \\ \hline
10& \begin{tabular}{@{}l@{}} Min \\ Latency   \end{tabular}  & Inception &ResNet152  
&219.6 &4.5 
&\textbf{178.2} & 5.6 
& 223.1 &4.5
& 202.3 & 5.2 
&\scriptsize{\begin{tabular}{@{}l@{}}17 \\ 135\end{tabular}} 
& \scriptsize{\begin{tabular}{@{}l@{}}DtoG \\ GtoD\end{tabular}} 
& \textbf{155.3} &6.4 
& 15 &15 \\ \hline
\end{tabular}
\label{tab:multiple_nn_results}
\end{table*}

\paragraph{\textit{Scenario 1 - Concurrent image processing with same DNNs:}} 
In systems aiming for high throughput, multiple instances of the same DNN could process consecutive images concurrently.
Fig~\ref{fig:single_nn_results} reports the results of five different experiments designed for this Scenario. The experiments are run on NVIDIA Orin, and we compare \sysname{} against two naïve baselines and Mensa~\cite{boroumand2021google}. 
Overall, our experiments for this scenario show that \sysname{} can boost throughput (\ie{} \textit{FPS}) up to 29\%. There are several key observations we make in this experiment: (1) In GoogleNet experiment, \sysname{} maps  the initial and middle groups of layers (1-95 and 38-149) to GPU for both DNN instances since those layers GPU is ~$\sim$2x faster than compared to DLA. 
(2) Non-collaborative GPU \& DLA execution does not always generate a better throughput compared to GPU-only execution due to shared memory contention. 
(3) We observe either limited improvements or no improvement by Mensa as it doesn't consider shared memory contention, leading to mismatched layer transitions. Even though Mensa considers transition costs, its greedy strategy fails to account for the transition costs occurring in the future, leading to inefficient transition decisions. 


\subsection{Concurrently Running Different Type of DNNs}\label{sec:multi_dnn}

Table~\ref{tab:multiple_nn_results} lists the results of the experiments we performed by comparing \sysname{} to naïve and state-of-the-art multi-DNN concurrent execution schemes. Experiments 1-5 are on Xavier AGX, 6-8 are on AGX Orin, and 9-10 are on Qualcomm 865. The second to fourth columns describe the experiment designs and the corresponding scenarios being run. There are four baselines we compare our work against: (1) \textit{GPU-only}; (2) the \textit{GPU\&DSA}; (3) and (4) \textit{Herald}~\cite{kwon2021hpca} and \textit{H2H}~\cite{h2h_dac}, which we identify as the closest and most relevant recent work. The last three columns list the optimal schedule found by \sysname{}, the latency and throughput (\ie{} FPS) for \sysname{}, and the improvement over the best performing baseline.

\paragraph{\textit{Scenario 2 - Two different DNNs operating on same data:}} This scenario illustrates a case where different DNNs, such as object detection and image segmentation, process the same input in parallel, and they synchronize afterward. The results are assumed to be passed on to subsequent workloads, such as motion planning~\cite{novickis2020functional}, and the loop is started over. Experiments 1, 2, 6, and 10 in Table~\ref{tab:multiple_nn_results} are designed for this scenario on three different target architectures.
Our results show that \sysname{} improves both latency and throughput up to 23\% in all four experiments of this scenario. 
We also observe that both H2H and Herald make inaccurate latency estimations that are wrong by up to 75\% since neither of them considers shared memory contention. While experiments 1 and 6 correspond to the execution of the same scenario but on different SoCs, \ie{} Xavier AGX and AGX Orin, \sysname{} results in different schedules for each device. For experiment 10, (2) GPU \& DSP is the best performing baseline for Qualcomm platform since GPU \& DSP are more balanced in terms of their computation capability in this platform. Even though the schedule found by \sysname{} in experiment 10 on Qualcomm has a relatively higher transition cost among other transition candidates, the improvement primarily comes from minimizing the memory contention and effectively distributing the layers to DSAs.  

\vspace{-0.6em}
\paragraph{\textit{Scenario 3 - Two different DNNs operating on streaming data:}} This scenario examines a common autonomous system setup where the input (\eg{} camera stream) is available as a data stream and multiple operations, such as object detection followed by object tracking~\cite{ravindran2020multi}, are applied in a pipelined manner. This scenario is covered in experiments 3, 4, 7, and 9. In order to establish the dependency among DNNs, we connect the last layer of the $DNN_1$ to the first layer of $DNN_2$ as an input. Interestingly, \sysname{} opts not to use DLA for any layer in scenario 4 since running two images sequentially on GPU yields in higher throughput in all cases. Assigning the second image to the DLA would increase the cumulative latency, regardless of whether the scheme is collaborative or not. So, if there are cases where layer-level mapping does not foster any benefits, \sysname{} is capable of identifying these cases and utilizing the baseline solution instead. This scheme guarantees no worse results over baselines.

\vspace{-0.6em}
\paragraph{\textit{Scenario 4 - Multiple DNNs with concurrent and streaming data:}} In this scenario, two DNNs ($DNN_1$ and $DNN_2$) have a serial dependency in between and another DNN ($DNN_3$) runs in parallel with the former two~\cite{ravindran2020multi}. 
Experiments 5 and 8 are designed for this scenario and the objective function is to minimize the combined latency. 
\sysname{} is able to provide latency and throughput improvements up to 22\%. 
Best performing baselines run $DNN_3$ mostly on GPU since unbalanced workloads among accelerators and shared memory contention alleviate the advantages of concurrent utilization. In Exp. 5, the schedules that are using both DSAs concurrently perform worse than serialized GPU executions, since DLA is generally less effective in running fully-connected layers. In Exp. 8, H2H provides the fastest baseline performance since they are capable of exploiting heterogeneity of DSAs for appropriate layers (\eg{} such as running DLA-efficient small layers on DLA whereas assigning others to the GPU). However, the schedule proposed by H2H leads to  over-subscribed DLA execution. On the other hand, \sysname{} finds the right transition point where no accelerators are overloaded and the transition cost is lower. In some schedules generated by H2H, such as Exp. 3, while the transition points that are identified by H2H prevent accelerator over-subscription, the assignments for the remaining layers on both DNNs lead to workload imbalance.

\begin{figure}
\includegraphics[width=0.95\linewidth]{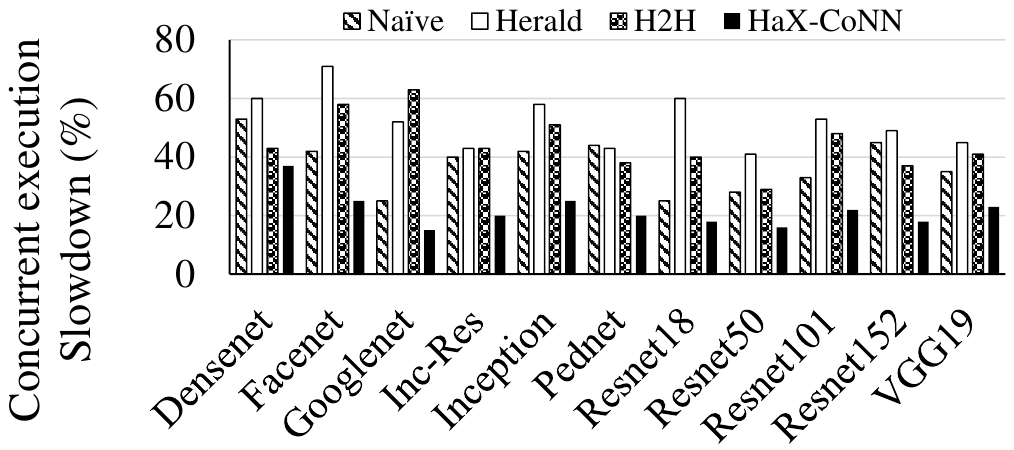}\\
\vspace{-0.8em}
\caption{Slowdown of concurrently executing GoogleNet on GPU with different DNNs on DLA.}
\vspace{-1.7em}
\label{fig:multi_dnn_slowdown}
\end{figure}

Overall, we observe that the benefits of architectural heterogeneity exploited by the state-of-the-art are limited. The primary reason for the subpar performance of H2H and Herald compared to naive baselines is that their cost functions ignore shared memory contention. This, in turn, causes the timings to be mispredicted and eventually results in being unable to generate optimal schedules. In those schedules, certain layers end up being assigned to the same accelerator (either GPU or DSA) at the same time and this is due to poor (\ie{} non-optimal) handling of constraints triggered by mispredicted execution times. 
For example, on the DLA, two layer groups that are supposed to execute at different times are scheduled together, but they end up waiting for each other. During this time the other accelerator (\eg{} the GPU) is left idle.

In experiments 4 and 9 of Table~\ref{tab:multiple_nn_results}, the latency of schedules generated by Herald is better than H2H because H2H makes optimizations to reduce the transition costs, yet leading to worse inter-DSA contention. We also observe that some of the optimizations performed by H2H are already performed by TensorRT 
Therefore, such optimizations are already included covered by our baselines, and this may hinder the benefits of H2H over our baselines. On the other hand, this situation does not affect the benefits demonstrated by \sysname{} over H2H and Herald. 
Also, it is worth noting it takes more time to generate schedules with H2H or Herald (\ie{} more than 10 seconds in most cases) than with HaX-CoNN.


Based on what we observe in Table ~\ref{tab:multiple_nn_results}, we further analyze the slowdown caused by memory contention. Fig.~\ref{fig:multi_dnn_slowdown} depicts the amount of
slowdown experienced by GoogleNet running on GPU when other DNNs are concurrently run on the DLA of Xavier AGX. 
The slowdown is calculated based on the standalone GPU execution of GoogleNet where there are no other concurrently running DNNs.  
\sysname{} significantly  reduces the shared memory contention slowdown in all experiments.

\subsection{Adapting Optimal Scheduling to Dynamically Changing Workloads}\label{sec:dynamic}


In our experiments, we observe that Z3 solver may take a few seconds to find the optimal schedules when running on a single CPU core of NVIDIA AGX Orin. Even though autonomous loops run continuously for extended periods, when CFG changes, stalling the computation for a few seconds to generate a new schedule may not be practical. As discussed in Section 2.5, the system must respond rapidly to dynamic changes in the environment.

To handle dynamic changes to the autonomous CFG, we propose D-\sysname{}, which operates as follows: (1) It starts with an initial best naïve schedule.\footnote[1]{ We do not start with a Herald or H2H schedule since they also take seconds to return a schedule.} (2) As the autonomous loop starts executing with the initial schedule, we periodically replace the initial schedule with a better schedule as Z3 progresses. Z3 deploys a method called \textit{model-based quantifier instantiation}, which works by modifying the candidate solution and evaluating the quantifier by back-forwarding the feedback to the model~\cite{marques2003grasp}. This helps Z3 to eliminate unsatisfiable solutions in the early stages of the execution, leading to significant improvements during the first few iterations. 
(3) We continue running Z3 until no further improvement is possible.

\begin{figure}[t]
    \centering
    \includegraphics[width=1\linewidth]{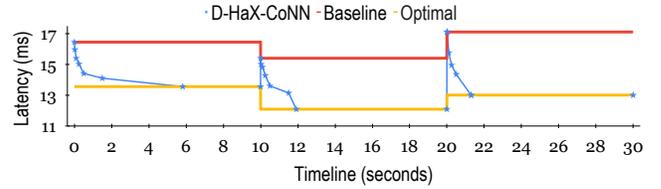}
    \vspace{-1.6em}
    \caption{A dynamic execution scenario where the target CFG (\ie{} DNN-pairs) changes every 10 seconds. D-\sysname{} is shown to gradually improve the execution time as Z3 is asked to update schedules at 25ms, 100ms, 250ms, 500ms, and 1.5s. Blue stars show the update intervals.}
    \label{fig:dynamic_haxconn}
    \vspace{-1.3em}
\end{figure}

To demonstrate the effectiveness of D-\sysname{}, we perform an experiment where dynamic changes in the CFG are simulated by changing three DNN-pairs being executed every 10 seconds. DNN pairs are the same within Exp. 2, 5, and 1 in Table~\ref{tab:multiple_nn_results}, respectively. 
Fig.~\ref{fig:dynamic_haxconn} depicts the concurrent execution time of the DNN-pairs (\ie{} latency per image) as they change. In this experiment, D-\sysname{} is run on a single CPU core with an initial schedule given by baseline. We update the schedules at 25ms, 100ms, 250ms, 500ms, and 1.5s after starting Z3. 
The blue lines correspond to the execution time of the updated schedule. The optimal schedule for each pair (represented by a yellow line) is statically calculated to denote the \textit{oracle} solution that D-\sysname{} is expected to reach.

Our results show that D-\sysname{} quickly converges to the optimal solution.  In particular, D-\sysname{} reaches an optimal solution faster for the second and third DNN-pairs (1.9s and 1.3), compared to the first pair (5.8s), since the latter pair has three DNNs and more layer groups. As explained before, a larger number of layer groups results in potentially more transition points to explore, which then increases the time required to explore all transition candidates for the optimized objective function. 

\begin{table}[b]
\scriptsize
\vspace{-1.5em}
\caption{The scheduling overhead (\%) of dynamically running the Z3 solver on a CPU core while AlexNet on the DLA is concurrently executed along with other DNNs on the GPU of Xavier Orin.}
\vspace{-1em}
\begin{tabular}{|c|c|c|c|c|c|}\hline
 CaffeNet & DenseNet & GoogleNet & Inc-res-v2 & Inception &MobileNet \\\hline  0.45\% & 0.89\% & 1.64\% & 0.69\%  & 1.64\% & 1.31\% \\\hline  ResNet18 & ResNet52 & ResNet101 & ResNet152 & VGG16 & VGG19 \\  \hline 
 0.16\% & \%0.23\% & 0.38\% & 0.71\% & 1.12\% & 1.59\% \\  \hline
\end{tabular}
\label{table:overhead}
\end{table}

To evaluate the overhead of running Z3 solver along with the concurrent DNN execution, we conduct another experiment where we run AlexNet on DLA along with various DNNs on GPU while Z3 solver runs on a single CPU core of NVIDIA AGX Orin. The results, presented in Table ~\ref{table:overhead}, show that running the solver on the fly slows down the DNN execution time by less than 2\%. This is primarily attributed to Z3's low memory footprint and Z3 converges the size of the parameter search space for our targeted problem into a smaller set.


\subsection{Exhaustive Evaluation with All DNN-pairs}
\label{sec:generalized_multi_dnn_experiments}
\vspace{-0.4em}

\begin{table}[t]
\caption{\normalsize{Comparison among \sysname{} and the best baseline for DNN pairs running on AGX Orin } }
\vspace{-0.3cm}
\footnotesize       
\begin{tabular}{|m{4.6em}|m{1.2em}|m{1.1em}|m{1.1em}|m{1.1em}|m{1.1em}|m{1.1em}|m{1.1em}|m{1.1em}|m{1.1em}|m{1.1em}|}\hline
\textbf{DNNs}
\scriptsize
&1
&2
&3
&4
&5
&6
&7
&8
&9
&10\\ \cline{1-11}

\scriptsize{1-CaffeNet}
&\begin{tabular}{@{}c@{}}\scriptsize{GPU} \\ \scriptsize{\textbf{\textcolor{darkGreen}{1.13}}}\end{tabular}  &&&&&&&&& \\ \hline

\scriptsize{2-DenseNet}
&\begin{tabular}{@{}c@{}}\scriptsize{GPU} \\ \scriptsize{\textbf{\textcolor{darkGreen}{1.14}}}\end{tabular}
&\begin{tabular}{@{}c@{}}\scriptsize{H2H} \\ \scriptsize{\textbf{\textcolor{darkGreen}{1.18}}}\end{tabular}  &&&&&&&&  \\ \hline

\scriptsize{3-GoogleNet}
&\begin{tabular}{@{}c@{}}\scriptsize{GPU} \\ \scriptsize{\textbf{\textcolor{darkGreen}{1.06}}}\end{tabular}
&\begin{tabular}{@{}c@{}}\scriptsize{D/G} \\ \scriptsize{\textbf{\textcolor{darkGreen}{1.18}}}\end{tabular}
&\begin{tabular}{@{}c@{}}\scriptsize{GPU} \\ \scriptsize{\textbf{\textcolor{darkGreen}{1.22}}}\end{tabular}  &&&&&&&  \\ \hline

\scriptsize{4-Inc-res-v2}
&\begin{tabular}{@{}c@{}}\scriptsize{GPU} \\ \scriptsize{\textbf{\textcolor{darkGreen}{1.08}}}\end{tabular}
&\begin{tabular}{@{}c@{}}\scriptsize{GPU} \\ \scriptsize{\textbf{x}}\end{tabular}
&\begin{tabular}{@{}c@{}}\scriptsize{D/G} \\ \scriptsize{\textbf{\textcolor{darkGreen}{1.25}}}\end{tabular}
&\begin{tabular}{@{}c@{}}\scriptsize{D/G} \\ \scriptsize{\textbf{\textcolor{darkGreen}{1.18}}}\end{tabular}  &&&&&& \\ \hline

\scriptsize{5-Inception}
&\begin{tabular}{@{}c@{}}\scriptsize{GPU} \\ \scriptsize{\textbf{\textcolor{darkGreen}{1.10}}}\end{tabular}
&\begin{tabular}{@{}c@{}}\scriptsize{GPU} \\ \scriptsize{\textbf{\textcolor{darkGreen}{1.15}}}\end{tabular}
&\begin{tabular}{@{}c@{}}\scriptsize{GPU} \\ \scriptsize{\textbf{\textcolor{darkGreen}{1.15}}}\end{tabular}
&\begin{tabular}{@{}c@{}}\scriptsize{D/G} \\ \scriptsize{\textbf{\textcolor{darkGreen}{1.06}}}\end{tabular}
&\begin{tabular}{@{}c@{}}\scriptsize{H2H} \\ \scriptsize{\textbf{\textcolor{darkGreen}{1.05}}}\end{tabular}  &&&&& \\ \hline

\scriptsize{6-ResNet18}
&\begin{tabular}{@{}c@{}}\scriptsize{GPU} \\ \scriptsize{\textbf{x}}\end{tabular}
&\begin{tabular}{@{}c@{}}\scriptsize{D/G} \\ \scriptsize{\textbf{\textcolor{darkGreen}{1.14}}}\end{tabular}
&\begin{tabular}{@{}c@{}}\scriptsize{G/D} \\ \scriptsize{\textbf{\textcolor{darkGreen}{1.13}}}\end{tabular}
&\begin{tabular}{@{}c@{}}\scriptsize{D/G} \\ \scriptsize{\textbf{\textcolor{darkGreen}{1.32}}}\end{tabular}
&\begin{tabular}{@{}c@{}}\scriptsize{GPU} \\ \scriptsize{\textbf{\textcolor{darkGreen}{1.19}}}\end{tabular}
&\begin{tabular}{@{}c@{}}\scriptsize{GPU} \\ \scriptsize{\textbf{\textcolor{darkGreen}{1.23}}}\end{tabular}  &&&& \\ \hline

\scriptsize{7-ResNet50}
&\begin{tabular}{@{}c@{}}\scriptsize{GPU} \\ \scriptsize{\textbf{x}}\end{tabular}
&\begin{tabular}{@{}c@{}}\scriptsize{D/G} \\ \scriptsize{\textbf{\textcolor{darkGreen}{1.21}}}\end{tabular}
&\begin{tabular}{@{}c@{}}\scriptsize{H2H} \\ \scriptsize{\textbf{\textcolor{darkGreen}{1.06}}}\end{tabular}
&\begin{tabular}{@{}c@{}}\scriptsize{GPU} \\ \scriptsize{\textbf{\textcolor{darkGreen}{1.16}}}\end{tabular}
&\begin{tabular}{@{}c@{}}\scriptsize{GPU} \\ \scriptsize{\textbf{\textcolor{darkGreen}{1.11}}}\end{tabular}
&\begin{tabular}{@{}c@{}}\scriptsize{GPU} \\ \scriptsize{\textbf{\textcolor{darkGreen}{1.06}}}\end{tabular}
&\begin{tabular}{@{}c@{}}\scriptsize{GPU} \\ \scriptsize{\textbf{\textcolor{darkGreen}{1.17}}}\end{tabular}  &&& \\ \hline

\scriptsize{8-ResNet101}
&\begin{tabular}{@{}c@{}}\scriptsize{GPU} \\ \scriptsize{\textbf{\textcolor{darkGreen}{1.11}}}\end{tabular}
&\begin{tabular}{@{}c@{}}\scriptsize{G/D} \\ \scriptsize{\textbf{\textcolor{darkGreen}{1.05}}}\end{tabular}
&\begin{tabular}{@{}c@{}}\scriptsize{G/D} \\ \scriptsize{\textbf{\textcolor{darkGreen}{1.08}}}\end{tabular}
&\begin{tabular}{@{}c@{}}\scriptsize{D/G} \\ \scriptsize{\textbf{\textcolor{darkGreen}{1.19}}}\end{tabular}
&\begin{tabular}{@{}c@{}}\scriptsize{GPU} \\ \scriptsize{\textbf{\textcolor{darkGreen}{1.08}}}\end{tabular}
&\begin{tabular}{@{}c@{}}\scriptsize{D/G} \\ \scriptsize{\textbf{\textcolor{darkGreen}{1.24}}}\end{tabular}
&\begin{tabular}{@{}c@{}}\scriptsize{GPU} \\ \scriptsize{\textbf{\textcolor{darkGreen}{1.11}}}\end{tabular}
&\begin{tabular}{@{}c@{}}\scriptsize{GPU} \\ \scriptsize{\textbf{\textcolor{darkGreen}{1.09}}}\end{tabular}  && \\ \hline

\scriptsize{9-ResNet152}
&\begin{tabular}{@{}c@{}}\scriptsize{GPU} \\ \scriptsize{\textbf{\textcolor{darkGreen}{1.09}}}\end{tabular}
&\begin{tabular}{@{}c@{}}\scriptsize{G/D} \\ \scriptsize{\textbf{\textcolor{darkGreen}{1.08}}}\end{tabular}
&\begin{tabular}{@{}c@{}}\scriptsize{G/D} \\ \scriptsize{\textbf{\textcolor{darkGreen}{1.17}}}\end{tabular}
&\begin{tabular}{@{}c@{}}\scriptsize{GPU} \\ \scriptsize{\textbf{\textcolor{darkGreen}{1.14}}}\end{tabular}
&\begin{tabular}{@{}c@{}}\scriptsize{Her.} \\ \scriptsize{\textbf{\textcolor{darkGreen}{1.07}}}\end{tabular}
&\begin{tabular}{@{}c@{}}\scriptsize{D/G} \\ \scriptsize{\textbf{\textcolor{darkGreen}{1.18}}}\end{tabular}
&\begin{tabular}{@{}c@{}}\scriptsize{H2H} \\ \scriptsize{\textbf{\textcolor{darkGreen}{1.09}}}\end{tabular}
&\begin{tabular}{@{}c@{}}\scriptsize{GPU} \\ \scriptsize{\textbf{\textcolor{darkGreen}{1.08}}}\end{tabular}
&\begin{tabular}{@{}c@{}}\scriptsize{GPU} \\ \scriptsize{\textbf{\textcolor{darkGreen}{1.18}}}\end{tabular}  & \\ \hline

\scriptsize{10-VGG19}
&\begin{tabular}{@{}c@{}}\scriptsize{GPU} \\ \scriptsize{\textbf{x}}\end{tabular}
&\begin{tabular}{@{}c@{}}\scriptsize{GPU} \\ \scriptsize{\textbf{\textcolor{darkGreen}{1.11}}}\end{tabular}
&\begin{tabular}{@{}c@{}}\scriptsize{GPU} \\ \scriptsize{\textbf{\textcolor{darkGreen}{1.04}}}\end{tabular}
&\begin{tabular}{@{}c@{}}\scriptsize{GPU} \\ \scriptsize{\textbf{x}}\end{tabular}
&\begin{tabular}{@{}c@{}}\scriptsize{GPU} \\ \scriptsize{\textbf{x}}\end{tabular}
&\begin{tabular}{@{}c@{}}\scriptsize{GPU} \\ \scriptsize{\textbf{\textcolor{darkGreen}{1.08}}}\end{tabular}
&\begin{tabular}{@{}c@{}}\scriptsize{GPU} \\ \scriptsize{\textbf{x}}\end{tabular}
&\begin{tabular}{@{}c@{}}\scriptsize{GPU} \\ \scriptsize{\textbf{x}}\end{tabular}
&\begin{tabular}{@{}c@{}}\scriptsize{GPU} \\ \scriptsize{\textbf{x}}\end{tabular}
&\begin{tabular}{@{}c@{}}\scriptsize{GPU} \\ \scriptsize{\textbf{x}}\end{tabular}   \\ \hline

\end{tabular}
\vspace{-0.4cm}
\label{tab:combination_nn_results}
\end{table}
\normalsize

The DNNs that are run concurrently in the experiments presented in Section~\ref{sec:multi_dnn} were handpicked to reflect the importance of the use cases in each scenario. In this subsection, we conduct a comprehensive evaluation of \sysname{}, by running every possible DNN-pair in our entire DNN set. Since we test every possible pair, the execution times for two concurrent DNNs can significantly differ. Therefore, for each pair of DNN, we first check the execution time on DLA and GPU for DNN-1 and compare it to DNN-2. 
Then, to balance out the discrepancy, we increase the number of iterations for the faster DNN. Such scenarios are quite common in multi-sensor systems where two independent sensor data (\ie{} camera and radar) are processed concurrently at different frequencies, or where multiple iterations over consecutive data are required to maintain the system's overall accuracy above a threshold. 
Results of this experiment are given in Table~\ref{tab:combination_nn_results} as a lower triangular matrix --The upper triangular matrix is symmetric because we are running DNN pairs.   
The first row of each cell shows the accelerator(s) where the baseline is the fastest for the corresponding objectives.
The second row of each cell shows the percentage of improvements that \sysname{} was able to achieve over the baseline. 
In this experiment, due to the complexity of the scheduling and because of similar reasons explained in Sec~\ref{sec:multi_dnn}, both H2H and Herald mostly result in worse runtimes than the naïve baselines. Key observations from this experiment include:
\begin{enumerate}
    \item Any pair involving GoogleNet shows improvement since GPU's performance is close to DLA's performance on GoogleNet and \sysname{} can exploit different transition points where both accelerators are efficient. 
    \item Overall, \sysname{} improves the throughput on 35 pairs out of 45 and identifies that GPU-only execution should be applied to the remaining 10 pairs, ensuring that \sysname{} does not underperform. However, experiments involving VGG19 show improvement only in three pairs. The fastest baselines for this DNN are all GPU-only and the execution of VGG19 on DLA is substantially slower than on GPU. Running another DNN on the DLA slows down the entire execution due to high memory contention. When DenseNet and GoogleNet are paired with VGG-19, \sysname{} shows a slight speedup since DLA is proportionally faster than the average on last layer groups in DenseNet and GoogleNet, and in the initial groups of VGG-19.
    \item Despite the execution of CaffeNet on DLA being considerably slower compared to GPU, which is a situation that favors GPU-only baseline, \sysname{} is still able to improve performance since CaffeNet is a compute-intensive DNN and does not cause too much contention when paired with other DNNs.
    
\end{enumerate}

\section{Conclusion}
We propose \sysname{}, a scheme that maps layers in concurrently executing DNN inference workloads to the accelerators of a heterogeneous SoC. \sysname{} holistically considers per-layer execution characteristics, shared memory contention, and inter-accelerator transitions while finding optimal schedules. Our experimental results show that \sysname{} can improve latency up to 32\%.

\bibliographystyle{plain}
\bibliography{references}

\end{document}